\documentclass{ws-ijmpa}

\def\GRG{{\it Gen. Relat. Gravit.} }

\def\MPL{{\it Mod. Phys. Lett.} }

\def\PL{{\it Phys. Lett.} }
\def\PR{{\it Phys. Rev.} }
\def\PRL{{\it Phys. Rev. Lett.} }

\def\PTP{{\it Progr. Theor. Phys.} }

\def\frac#1#2{{\textstyle{{#1}\over {#2}}}}

\def\lsim{\mathrel{\rlap{\lower4pt\hbox{\hskip1pt$\sim$}}
    \raise1pt\hbox{$<$}}}
\def\gsim{\mathrel{\rlap{\lower4pt\hbox{\hskip1pt$\sim$}}
    \raise1pt\hbox{$>$}}}
\def\sqr#1#2{{\vcenter{\vbox{\hrule height.#2pt
         \hbox{\vrule width.#2pt height#1pt \kern#1pt
         \vrule width.#2pt}
         \hrule height.#2pt}}}}

\def\beq{\begin{equation}}
\def\eeq{\end{equation}}
\def\beqa{\begin{eqnarray}} 
\def\eeqa{\end{eqnarray}}

\def\laq{\raise 0.4 ex \hbox{$<$}\kern -0.8 em\lower 0.62 ex\hbox{$\sim$}}
\def\gaq{\raise 0.4 ex \hbox{$>$}\kern -0.7 em\lower 0.62 ex\hbox{$\sim$}}

\begin{document}

\markboth{Orfeu Bertolami}
{Noncommutative scalar field minimally coupled to gravity}

%
\catchline{}{}{}{}{}
%

\title{Noncommutative scalar field minimally coupled to gravity\footnote{Talk presented at the Workshop on Quantum Gravity and 
Noncommutative Geometry, 20-23 July 2004, Universidade Lus\'ofona, Lisbon, Portugal.}}

\author{Orfeu Bertolami\footnote{E-mail: orfeu@cosmos.ist.utl.pt}}

\address{Instituto Superior T\'ecnico, Departamento de F\'\i sica \\
Av. Rovisco Pais 1, Lisbon, 1049-001, Portugal}




\maketitle


\begin{abstract}
A model for noncommutative scalar fields coupled to gravity based on the 
generalization of the Moyal product is proposed. Solutions compatible with homogeneous and isotropic 
flat Robertson-Walker spaces to first non-trivial order in the perturbation of the star-product 
are presented. It is shown that in the context of a typical chaotic inflationary scenario, 
at least in the slow-roll regime, noncommutativity yields no observable effect. 

\end{abstract}

\ccode{PACS numbers: 98.80-Cq, 04.50+h}

\vspace{0.5cm}

\centerline{{\bf Dedicated to the memory of Lu\'\i s Guisado}}

\vspace{0.5cm}

\section{Introduction}

The idea of noncommuting spatial coordinates is actually quite old and has been suggested by Snyder\cite{Snyder} 
about the time Quantum Field Theory itself was emerging as a consistent description of the fundamental interactions. 
More recently, noncommutative geometry has been systematized
by Connes\cite{Connes} and Woronowicz\cite{Woronowicz},
via the generalized concept of differential structure of generic (\( C^{\star } \))-algebras.
This formulation has been proposed as a possible formulation for quantum gravity via noncommutative differential
calculus\cite{Varilly}. In another fundamental setting, it has been pointed
out by Seiberg and Witten\cite{SW}, that noncommutative geometry
arises in the context of string theory, which has naturally motivated a
great interest in the subject. 

This interest has led to the construction of noncommutative field theories 
through the Moyal deformation of the product
of functions, which defines a noncommutative algebra\cite{Szabo,DougNekra}. 
In this type of setup the issues of unitarity\cite{Bahns} and renormalizability
cannot be fully understood, as the resulting particle
physics models are regarded as effective theories, even though, to some
extent, interesting bounds on the magnitude of the noncommutative parameter can be obtained\cite{Hewett}.

In this contribution the implications of the generalization of the noncommutative
algebra for the multiplication of tensors that are minimally coupled to a classical gravity field\cite{Bertolami1} 
is studied. It is hoped that this noncommutative
algebra approach may provide some insight into the physics of the Planck scale.
As it will be seen, interestingly, this setting can be tested via its impact in inflationary
models and, hence, on the Gaussian character of energy density fluctuations
or on the isotropy of the observables. Noncommutativity of coordinates
introduces a new fundamental length scale whose imprint may turn out, thanks the intervention of inflation, 
to have some observational consequences\cite{Lizzi1,Green}. 
The approach suggested here is similar
to the study of Ref. [13], even though differences
in details lead to somewhat different conclusions. Most remarkably, it is 
found that within a perturbation approach in an homogeneous and
isotropic background metric, the impact of noncommutativity in the context of 
the chaotic inflationary model\cite{Linde} is negligible.

Another fundamental issue that has been much discussed in the context of 
noncommutative field theories concerns the breaking of Lorentz 
invariance\cite{Carroll}. Actually, the possibility that this fundamental symmetry of Nature is 
broken has been widely discussed in the recent literature\cite{Kostelecky1}. 
Indeed, the spontaneous breaking of Lorentz symmetry 
may arise in string/M-theory due to 
non-trivial vacuum solutions in string field theory\cite{Kostelecky2}, 
in loop quantum gravity\cite{Gambini}, in quantum gravity inspired spacetime foam 
scenarios\cite{Garay}, or via the spacetime variation of fundamental coupling constants\cite{Lehnert}. 
The breaking of Lorentz symmetry can, at least in principle, be tested in studies of ultra-high energy cosmic rays\cite{Sato}. 

In this work it is shown that Lorentz 
invariance may hold at least at first non-trivial order in perturbation theory of the 
noncommutative parameter\cite{Bertolami1}. Actually, the idea that the noncommutative parameter may 
be a Lorentz tensor has been considered in some field theory models\cite{Bertolami2}.

The work in which this contribution is based has been developed 
in collaboration with Lu\'\i s Guisado. Lu\'\i s was tragically killed in a car accident 
on June 28th, 2003. He was a brilliant 23 years old graduate student and a hope 
of the young generation of Portuguese theoretical physicists. I dedicate this contribution to his memory.

\section{Generalized Moyal Product}

Noncommutativity in Minkowski can be introduced via the so-called noncommutative
Moyal product defined as 

\begin{equation}
 T*W\left( x\right) = \sum _{n=0}^{\infty }\frac{\left( i/2\right) ^{n}}{n!}\theta ^{\alpha _{1}\beta _{1}}\ldots \theta ^{\alpha _{n}\beta _{n}}\left( T_{,\alpha _{1}\ldots \alpha _{n}}\right) \left( W_{,\beta _{1}\ldots \beta _{n}}\right) ~,
\end{equation}
where \( T \) and \( W \) are generic tensors whose indices have
been suppressed, the primes denote partial derivatives and \( \theta ^{\alpha \beta } \)
is often taken to be a constant. Aiming to preserve Lorentz symmetry to start with, we consider \( \theta ^{\alpha \beta } \)
as a spacetime dependent antisymmetric Lorentz tensor. Thus, the commutator between
coordinates is given by 

\begin{equation}
\label{commutator}
\left[ x^{\mu },x^{\nu }\right] =i\theta ^{\mu \nu }(x)~.
\end{equation}
Thus, our suggestion in order to preserve general covariance is to consider instead the following
generalized Moyal product 

\begin{equation}
 T*W\left( x\right) =
 \sum ^{\infty }_{n=0}\frac{\left( i/2\right) ^{n}}{n!}\theta ^{\alpha _{1}\beta _{1}}\ldots \theta ^{\alpha _{n}\beta _{n}}\left( T_{;\alpha _{1}\ldots \alpha _{n}}\right) \left( W_{;\beta _{1}\ldots \beta _{n}}\right) ~, 
\label{Moyal_gen}
\end{equation}
where the semicolon denotes covariant derivative with respect to the Levi-Civita
connection and \( \theta ^{\alpha \beta } \) is a non-constant rank-2
antisymmetric tensor. This proposal, despite of being non-associative in general, implies 
that this property may be recovered to some extent for a scalar field, \( \Phi  \), 
through the condition \( \theta ^{\alpha \beta }\Phi _{;\alpha }=0 \).

By use of the antisymmetry of \( \theta ^{\alpha \beta } \)
one can easily show that, under conjugation, \( \left( T*W\right) ^{*}=W^{*}*T^{*} \).
The compatibility of the metric yields \( g^{\mu \nu }*T=g^{\mu \nu }T \)
so that the operation of raising and lowering of indices is not affected
by noncommutativity.

Noncommutative Lagrangian densities are obtained by substituting the usual
products into star-products so that one has to evaluate integrals
of the form 

\begin{equation}
S=\int d^{4}x\sqrt{-g}~T^{*}*W ~.
\end{equation}

Integrating by parts and dropping surface terms,
one can arrange the covariant derivatives on the star-product to act
either on \( T \) or on \( W \), that is 

\begin{equation}
S =  \int d^{4}x\sqrt{-g}~T^{*}\left( {\mathcal{A}}W\right)  =  \int d^{4}x\sqrt{-g}\left( {\mathcal{A}}T\right) ^{*}W ~,
\end{equation}
where \( {\mathcal{A}} \) is an Hermitian operator given by

\begin{equation}
{\mathcal{A}}W= \sum _{n=0}^{\infty }\frac{\left( -i/2\right) ^{n}}{n!}
\left[ \theta ^{\alpha _{1}\beta _{1}}\ldots \theta ^{\alpha _{n}\beta _{n}}\left( W_{;\beta _{1}\ldots \beta _{n}}\right) \right] _{;\alpha _{n}\ldots \alpha _{1}} ~.
\end{equation}

In the case the Lagrangian density is quadratic on the tensor \( T \) one
can use the property under conjugation to demonstrate that \( T*T \) is
real, and therefore that

\begin{equation}
S'=  \int d^{4}x\sqrt{-g}~T*T=\int d^{4}x\sqrt{-g}T\left( {{\mathcal{A}}+{\mathcal{A}}^{*}\over 2}\right) T 
\equiv  \int d^{4}x\sqrt{-g}T{\mathcal{O}}T ~,
\end{equation}
where the Hermitian operator \( {\mathcal{O}} \equiv \frac{1}{2}\left( {\mathcal{A}}+{\mathcal{A}}^{*}\right)  \)
has been introduced

\begin{equation}
{\mathcal{O}}W= 
\sum _{n=0}^{\infty }\frac{\left( -1/4\right) ^{n}}{\left( 2n\right) !}\left[ \theta ^{\alpha _{1}\beta _{1}}\ldots \theta ^{\alpha _{2n}\beta _{2n}}\left( W_{;\beta _{1}\ldots \beta _{2n}}\right) \right] _{;\alpha _{2n}\ldots \alpha _{1}} ~. 
\label{op_o}
\end{equation}

\section{Noncommutative scalar field coupled to gravity}

\subsection{Massive scalar field}

The noncommutative action for a massive scalar field, \( \Phi  \),
is quadratic, and so, from the previous results 

\begin{equation}
S=-\frac{1}{2}\int d^{4}x\sqrt{-g}\left[ \nabla ^{\mu }\Phi {\mathcal{O}}\nabla _{\mu }\Phi +m^{2}\Phi {\mathcal{O}}\Phi \right] ~;
\end{equation}
the equation of motion being given by 

\begin{equation}
\label{motion_fi}
\nabla ^{\mu }{\mathcal{O}}\nabla _{\mu }\Phi -m^{2}{\mathcal{O}}\Phi =0 ~.
\end{equation}

The Hermitian operator \( {\mathcal{O}} \) naturally arises in the equations of
motion, corresponding 
to an observable of the scalar field. In the commutative
limit, \( lim_{\theta \rightarrow 0}{\mathcal{O}}=1 \). On the other
hand, switching off gravity and admitting that \( \theta ^{\alpha \beta } \)
is constant, yields \( {\mathcal{O}}=1 \), since the partial derivatives
commute and are contracted with the antisymmetric tensor \( \theta ^{\alpha \beta } \)
in Eq. (\ref{op_o}). Thus, in this model noncommutativity arises only through the coupling to gravity. 
This has its origin on the fact that the usual Moyal
product obeys, under integration, the cyclic property

\begin{equation}
\int d^{4}x\, f*g=\int d^{4}x\, f\, g=\int d^{4}x\, g*f ~.
\end{equation}

\subsection{Scalar field with an arbitrary potential}

We consider now the noncommutative generalization of an
arbitrary analytic commutative potential \( V\left( \Phi \right)  \).
Associativity played no role in the case of a massive scalar field because one dealt
with a quadratic action. Now, however, for an arbitrary potential, in general, 
the resulting star-product is not associative. 

Given a commutative analytic potential 

\begin{equation}
V\left( \Phi \right) =\sum ^{\infty }_{n=0}{\lambda _{n}\over n!}\Phi ^{n} ~,
\end{equation}
its corresponding noncommutative version has the form

\begin{equation}
V_{NC}\left( \Phi \right) =\sum _{n=0}^{\infty }{\lambda _{n}\over n!}\overbrace{\Phi *\ldots *\Phi }^{n\: factors} ~,
\end{equation}
provided the corresponding action of the star-product upon powers of the scalar
field is associative. This generalization is considered 
for the case where \( \theta ^{\alpha \beta }\Phi _{;\beta }=0 \).

Since there is no a priori associativity, let us consider the sequence

\begin{equation}
s_{2}=\left( \Phi *\Phi \right) \qquad s_{n+1}=\Phi *s_{n} ~,\; \; \; n>2 ~.
\end{equation}
It is not difficult to prove that, up to second order,

\begin{equation}
s_{n}\simeq \Phi ^{n}+{n\left( n-1\right) \over 2}\Phi ^{n-2}\left( \Phi \hat{*}\Phi \right) 
\end{equation}
where it is natural to define

\begin{equation}
\varphi \hat{*}\chi  \equiv -\frac{1}{8}\theta ^{\alpha _{1}\beta _{1}}\theta ^{\alpha _{2}\beta _{2}}\left( \varphi _{;\alpha _{1}\alpha _{2}}\right) \left( \chi _{;\beta _{1}\beta _{2}}\right)~.
\end{equation}
Moreover, for every \( m \) and \( n \), one can show that, up to second order,

\begin{equation}
s_{n}*s_{m}\simeq s_{m+n}~,
\end{equation}
which demonstrates that one can compute the power \( s_{q} \) grouping
\( q \) star-products in any combination one wishes. Therefore, the star-product
of powers of \( \Phi  \) turns out to be associative. These results allow writing

\begin{equation}
\label{ncpot}
V_{NC}\left( \Phi \right) \equiv V\left( \Phi \right) +\frac{1}{2}V''\left( \Phi \right) \left( \Phi \hat{*}\Phi \right)~,
\end{equation}
so that \( '=d/d\Phi  \). 

The variation of the potential \( V_{NC} \) in the action yields

\begin{equation}
-{\delta S_{pot}\over \delta \Phi }=V'+\frac{1}{2}V'''\left( \Phi \hat{*}\Phi \right) -\frac{1}{4}{\mathcal{F}}\left[ V,\Phi \right]~,
\end{equation}
 where the operator has been defined

\begin{equation}
{\mathcal{F}}\left[ V,\Phi \right]  \equiv \left[ \frac{1}{2}V''\theta ^{\alpha _{1}\beta _{1}}\theta ^{\alpha _{2}\beta _{2}}\phi _{;\beta _{1}\beta _{2}}\right] _{;\alpha _{2}\alpha _{1}}~.
\end{equation}

With this definition one also finds that 
\begin{equation}
{\mathcal{O}}\Phi _{;\mu }\simeq \Phi _{;\mu }-\frac{1}{8}{\mathcal{F}}\left[ \Phi ^{2},\Phi _{;\mu }\right]~.
\end{equation}

\subsection{Homogeneous and Isotropic Spacetime}

In what follows it is assumed that gravity is described by the Einstein-Hilbert action being 
therefore unaffected by noncommutativity. The aim of this proposal is to study
the impact of the noncommutative algebra of tensors on a non-trivial spacetime background. Furthermore, it 
should be pointed out that there is no canonical way of introducing the noncommutative algebra within the geometrical
formulation of gravity, namely in the Riemann tensor and, ultimately,
in the Ricci scalar. It the follows that the Einstein equations in the presence of a noncommutative scalar field 
are given by 

\begin{equation}
\label{motion_grav}
R_{\alpha \beta }=-8\pi k\left[ \frac{1}{2}\nabla _{\{\alpha }\Phi {\mathcal{O}}\nabla _{\beta \}}\Phi +g_{\alpha \beta }V_{NC}\left( \Phi \right) \right] .
\end{equation}

As a concrete model, we analyze a homogeneous and isotropic space-time
described by the spatially flat Robertson-Walker metric 

\begin{equation}
ds^{2}=-dt^{2}+R^{2}\left( t\right) \left( dx^{2}+dy^{2}+dz^{2}\right) ~,
\end{equation}
where \( R(t)\) is the scale factor.
The non-vanishing components of the Christoffel symbols are the following 

\begin{equation}
\Gamma _{\: ij}^{t}=R\dot{R}\delta _{ij}\qquad \Gamma _{\: jt}^{i}={\dot{R}\over R}\delta _{\: j}^{i}
\end{equation}
and, as is well known, the Ricci tensor is diagonal:

\begin{equation}
R_{tt}=3{\ddot{R}\over R}\qquad R_{ij}=-\left( R\ddot{R}+2\dot{R}^{2}\right) \delta _{ij} ~.
\end{equation}

The non-trivial components of the antisymmetric noncommutative tensor,
\( \theta ^{\alpha \beta } \), correspond to two 3-vectors which
we denote by \( \vec{E} \) and \( \vec{B} \), in analogy with the
electromagnetic tensor. The following notation is used: 

\begin{equation}
\label{theta_param}
\theta ^{\alpha \beta }=\left( \begin{array}{cccc}
0 & -E_{x} & -E_{y} & -E_{z}\\
E_{x} & 0 & -B_{z} & B_{y}\\
E_{y} & B_{z} & 0 & -B_{x}\\
E_{z} & -B_{y} & B_{x} & 0
\end{array}\right) .
\end{equation}

\noindent
It is relevant to point out that, even if \( \theta ^{\alpha \beta } \)
is homogeneous, \( \theta ^{\alpha \beta }=\theta ^{\alpha \beta }\left( t\right)  \),
it is still quite possible that symmetry under rotations is broken and some
attention should be paid concerning the choice of an isotropic Ansatz 
for the metric, as \( \vec{E} \)
and \( \vec{B} \) can give rise to preferred directions in space. It can be shown, 
however, that there is a noncommutative model consistent with
homogeneity and isotropy to first order in perturbation theory, for
the homogeneous scalar field, \( \partial _{i}\Phi =0 \). Under these
conditions, it follows from Eqs. (\ref{motion_fi}) and (\ref{motion_grav}) that

\begin{equation}
\ddot{\Phi }+3{\dot{R}\over R}\dot{\Phi }+V'= 
{\partial _{t}\left( R^{3}{\mathcal{F}}\left[ \Phi ^{2},\Phi _{;t}\right] \right) \over 8R^{3}}+\frac{1}{2}V'''
\left( \Phi \hat{*}\Phi \right) +\frac{1}{4}{\mathcal{F}}\left[ V,\Phi \right] ~,
\label{mov_pert_1}
\end{equation}

\begin{equation}
\left( {\dot{R}\over R}\right) ^{2}= 
{8\pi k\over 3}\left( \frac{1}{2}\dot{\Phi }^{2}+V+\frac{1}{2}V''\left( \Phi \hat{*}\Phi \right) -\frac{1}{16}\dot{\Phi }{\mathcal{F}}\left[ \Phi ^{2},\dot{\Phi }\right] \right) ~.
\label{mov_pert_2}
\end{equation}
The interested reader can find the explicit computation of
these terms in the Appendix of Ref. [10], the results being:

\begin{equation}
\label{nc_terms}
\begin{array}{c}
\Phi \hat{*}\Phi =-{1\over 2}\left( R\dot{R}\dot{\Phi }B\right) ^{2} ~,\\
{\mathcal{F}}\left[ V,\Phi \right] ={1\over 2R^{3}}\partial _{t}\left[ R^{5}\dot{R}^{2}\dot{\Phi }B^{2}\frac{1}{2}V''\right] ~,\\
{\mathcal{F}}\left[ \Phi ^{2},\dot{\Phi }\right] =-{2\over R^{3}}\partial _{t}\left[ R^{6}\dot{R}^{2}B^{2}\partial _{t}\left( {\dot{\Phi }\over R}\right) \right] ~,
\end{array}
\end{equation}
where the condition \( \vec{E}=0 \) \cite{Unit} has been used. This condition
ensures that \( \theta ^{\alpha \beta }\Phi _{;\beta }=0 \)
and that the noncommutative generalization of the scalar potential Eq. (\ref{ncpot})
makes sense. Hence we see that the dependence of Eqs. (\ref{nc_terms})
in \( \theta ^{\alpha \beta } \) occurs only via \( B^{2} \) and
consequently invariance under rotations is preserved. Since the dynamics of the \( \vec{B} \) field 
is unknown, we consider, the logical choice

\begin{equation}
B^{2}=\hat{B}^{2}R^{-2\varepsilon },
\end{equation}
where \( \hat{B}^{2} \) is a constant. The parameter \( \varepsilon  \) will be determined in the next section.

\section{Slow-roll in Chaotic Inflation}

Since the effects of noncommutativity are expected to manifest at high energies, it is 
quite natural to study its influence in the inflationary process. Given the generality of 
conditions for the onset of inflation,
chaotic models \cite{Linde} are particularly suited for studying
the effect of noncommutativity. We look for solutions of Eqs. (\ref{mov_pert_1}) and
(\ref{mov_pert_2}) in first order of perturbation theory in \( \hat{B}^{2} \), considering solutions 
of the following form 

\begin{equation}
\Phi =\phi +\hat{B}^{2}\varphi \qquad R=a+\hat{B}^{2}\chi ~,
\end{equation}
where \( \Phi  \) and \( a \) are solutions of the unperturbed
(commutative) problem, while \( \varphi  \) and \( \chi  \) are
arbitrary time dependent functions to be determined. We neglect in
every step higher order terms in \( \hat{B}^{2} \). Using units in
which \( k=1 \), Eqs. (\ref{mov_pert_1}) and (\ref{mov_pert_2}) assume
the form 

\begin{eqnarray}
&  & \ddot{\Phi }+3{\dot{R}\over R}\dot{\Phi }+V'=\hat{B}^{2}f ~,\\
&  & \left( {\dot{R}\over R}\right) ^{2}={8\pi \over 3}\left( \frac{1}{2}\dot{\Phi }^{2}+V\right) +
{8\pi \over 3}\hat{B}^{2}g\label{eqmotion} ~, 
\end{eqnarray}
in terms of functions \( f \) and \( g \) which are specified below. Standard
perturbation procedure yields the usual inflationary equations

\begin{eqnarray}
 &  & \ddot{\phi }+3{\dot{a}\over a}\dot{\phi }+V'\left( \phi \right) =0~,\label{slow_eqs} \\
 &  & \left( {\dot{a}\over a}\right) ^{2}={8\pi \over 3}\left[ {1\over 2}\dot{\phi }^{2}+V\left( \phi \right) \right] ~.\label{Fried} 
\end{eqnarray}

The onset of inflation and slow-roll regime are achieved
once the following conditions are satisfied

\begin{equation}
\label{slow_cond}
{V'\over V}\leq \sqrt{48\pi }\quad ,\quad {V''\over V}\leq 24\pi  ~,
\end{equation}
so that we can drop the term \( \ddot{\phi } \) in the Eq. (\ref{slow_eqs})
and the kinetic term of the scalar field in Eq. (\ref{Fried}). Hence, 
the useful condition arises

\begin{equation}
\label{maj_fipto}
\left| \dot{\phi }\right| \leq \sqrt{2}V^{1/2}.
\end{equation}

\noindent
It then follows that terms in Eqs. (\ref{nc_terms}) can be estimated
using the slow-roll conditions and one finds\cite{Bertolami1} that all of them are proportional to \( a^{4-2\varepsilon } \)
and to factors that depend on \( V \) and \( \dot{\phi } \). Naturally, since during inflation
the Universe is expanding exponentially, the perturbation
theory is meaningful only if \( \varepsilon \geq 2 \). However, if \( \varepsilon >2 \)
it implies that the terms in Eqs. (\ref{nc_terms}) decay so swiftly
that noncommutativity will have no impact. Therefore, it can be concluded
from the consistency of perturbation theory that \( \varepsilon =2 \). Notice, that
this is a quite natural choice from the theoretical point of view.
Indeed, most of the studied noncommutative models consider a constant \( \theta ^{\alpha \beta } \); thus 
requiring that this is so for the physical coordinates \( y^{i}=R\, x^{i} \),
then one finds, from Eq. (\ref{commutator}), \( \left[ y^{i},y^{j}\right] =\hat{B}^{ij} \), 
implying that \( \varepsilon =2 \).

The equations for the perturbed terms are obtained gathering
all terms proportional to \( \hat{B}^{2} \) and it follows that this constant cancels
out from the differential equations. Function \( \varphi  \) satisfies
the relationship

\begin{equation}
f-4\pi {\dot{\phi }\over \dot{a}/a}g= 
\ddot{\varphi }+3{\dot{a}\over a}\left[ 1+{4\pi \over 3}
\left( {\dot{\phi }\over \dot{a}/a}\right) ^{2}\right] \dot{\varphi }+
V''\left[ 1+4\pi {V'\over V''}{\dot{\phi }\over \dot{a}/a}\right] \varphi ~,
\label{eqvarphi}
\end{equation}
where functions \( f \) and \( g \) can be computed
using the slow-roll conditions\cite{Bertolami1}: 

\begin{eqnarray}
 &  & \left| f\right| \leq \frac{1}{2}a_{1}V^{2}V''+\frac{1}{2}a_{2}V^{2}V'''+a_{3}V^{3}+a_{4}V^{5/2},\nonumber \\
 &  & \left| g\right| \leq a_{5}V^{3}+\frac{1}{2}a_{6}V^{2}V'',
\end{eqnarray}
with \( a_{1}\simeq 85.5 \) , \( a_{2}\simeq a_{6}\simeq 4.2 \), 
\( a_{3}\simeq 3.30\times 10^{3} \) , \( a_{4}\simeq 4.52\times 10^{3} \)
and \( a_{5}\simeq 1.76\times 10^{2} \).

To further proceed, it should be reminded that potentials in chaotic inflation are characterized by
a small overall coupling constant, \( \lambda \simeq 10^{-14} \), so
to ensure consistency with the amplitude of energy density perturbations
around \( 10^{-5} \), for \( \phi  \) field values of a few Planck
units. Thus, writing the potential as 

\begin{equation}
\label{pot}
V\left( \Phi \right) =\lambda \, v\left( \Phi \right) ,
\end{equation}
and as 
\begin{equation}
\label{maj_pot}
v\leq 10^{2},
\end{equation}
it implies that \( \left| f\right| \leq 4.5\times 10^{-27} \)
and \( \left| g\right| \leq 1.8\times 10^{-34} \), while the second
and the third terms of the right-hand side of Eq. (\ref{eqvarphi}) are of the order
\( 2.5\times 10^{-6} \) and \( 7\times 10^{-11} \), respectively.
Hence, for numerical purposes, the left-hand side of the Eq. (\ref{eqvarphi})
is vanishingly small and in this case, one obtains essentially the
same differential equation that would arise when performing perturbation
theory on the standard slow-roll approximation with no extra physics.
The conclusion is that noncommutativity introduces no change
in inflationary slow-roll physics for the inflaton field in the context of the chaotic model.

Moreover, from the equation for the \( \chi  \) perturbation 

\begin{equation}
\label{eqchi}
{d\over dt}\left( {\chi \over a}\right) ={4\pi \over 3\dot{a}/a}\left( \dot{\phi }\dot{\varphi }+V'\varphi +g\right) 
\end{equation}
one finds that the upper limit for \( \left| g\right|  \) implies
that this equation is not changed as well.
Thus, one can conclude that the results of the perturbation approach indicate
that the noncommutative aspects of the proposed model yield no impact on the 
chaotic inflationary model.

\section{Conclusions}

In this contribution we have studied the physics of
a noncommutative scalar field coupled to gravity via an extension
of the Moyal product. The general features
of the formalism were developed and its application in the context of a spatially flat Robertson-Walker 
metric were obtained. Results were found through perturbation methods, 
which necessarily require that the antisymmetric noncommutative tensor,
\( \theta ^{\alpha \beta } \), is small compared to the covariant
derivative of the fields. It has been shown that although there exists no 
equation for \( \theta ^{\alpha \beta } \),
both perturbation theory and theoretical considerations, allow concluding that
\( \theta ^{\alpha \beta }\sim R^{-2} \), where \( R \) is the scale
factor.

The antisymmetric tensor \( \theta ^{\alpha \beta } \)
can be parameterized by two three-vectors, just like in the case of
the electromagnetic tensor (c.f. Eq. (\ref{theta_param})).
The homogeneity requirement, that is, \( \partial _{i}\theta ^{\alpha \beta }=0 \),
could still lead to preferred directions in space rendering 
the Robertson-Walker metric Ansatz meaningless. Nevertheless, it is shown that, at least in first
order in perturbation theory, that does not occur since the terms arising
from noncommutative contributions depend only on the rotationally
invariants \( E^{2} \) and \( B^{2} \).

Furthermore, in the context of the slow-roll regime of a typical
chaotic inflation, it is shown that noncommutativity introduces negligible
effects. This is due mainly to two reasons. First, the scale parameter 
does not appear in the first order
terms as \( \theta ^{\alpha \beta }\sim R^{-2} \), otherwise these
would grow exponentially rendering perturbation theory meaningless.
On the other hand, the slow-roll conditions induce small derivative terms
for the inflaton field, Eq. (\ref{maj_fipto}), and for the logarithm
of the scale factor, Eq. (\ref{Fried}). Since the Moyal product is highly non-local 
as it involves many derivatives, the smallness of the noncommutative
contributions is a natural implication. In other words: as perturbation theory requires that \( \theta ^{\alpha \beta } \)
is small compared to the the derivative terms and these are themselves quite small. Thus, one is led
to conclude that noncommutative effects, if any, must arise
beyond the perturbation regime.

In summary, one can say that the present calculations
assume that perturbation theory is valid from a given cosmological
time \( t_{*} \) onward; thus, if the conditions for inflation are
met and \( B=\hat{B}R^{-2} \), then noncommutativity
has no impact in the chaotic inflationary scenario. 
This implies that \( B_{*}=\hat{B}R_{*}^{-2}\ll 1 \),
or \( \hat{B}\ll R^{2}_{*} \), and therefore, a small \( \hat{B} \) 
ensures the validity of perturbation theory
for any given \( R_{*} \). It is important to realize that the constant \( \hat{B} \) cancels
out in the perturbed differential equations, so its magnitude plays
no role on the smallness of the extra terms in Eqs. (\ref{eqvarphi}) and (\ref{eqchi}). These terms, on their turn, are small
as they involve high-order derivatives of the
scalar potential which has a small coupling constant.

Prior to \( t_{*} \), no model for \( B \) is proposed. Actually, even if the expression
\( B=\hat{B}R^{-2} \) or any other one with a singularity for \( B \)
at \( R=0 \) holds, this would occur before perturbation theory
is valid. However, if \( t_{*} \) coincides with the onset of
inflation, then the physics prior to \( t_{*} \) has negligible impact,
as chaotic initial conditions are satisfied.

There is also another scenario in which these considerations might
remain valid. If beyond perturbation effects allow for inflation, then
it is feasible that initially inflation is driven by noncommutativity 
and, at a later time, by the mechanism discussed here.

\section{Acknowledgments}

It is a pleasure to thank the colleagues Nuno Dias, Aleksandar Mikovic and Jo\~ao Prata 
as well as the Departamento de Matem\'atica e Ci\^encias da Computa\c c\~ao da 
Universidade Lus\'ofona 
for the organization of the Workshop on Quantum Gravity and 
Noncommutative Geometry, a most interesting and pleasant meeting.


\begin{thebibliography}{10}
\bibitem{Unit} Some authors suggest that unitarity requires \( \vec{E}=0 \) (see e.g. [13] and references therein).
\end{thebibliography}
\end{document}